\documentstyle[preprint,aps,eqsecnum]{revtex}
\begin{document}
\tighten
\draft
\title{INFLATON DECAY IN DE SITTER SPACETIME}
\maketitle
\begin{center}
{\it{\bf D. Boyanovsky$^{(a)}$, R. Holman$^{(b)}$, S. Prem Kumar$^{(b)}$}}
\end{center}
\begin{center}
{\it{(a) Department of Physics and Astronomy, University of Pittsburgh,
PA.\\15260,U.S.A.}}
\end{center}
\begin{center}  
{\it{(b) Department of Physics, Carnegie Mellon
University, Pittsburgh, PA.15213, U.S.A.}}
\end{center}

\begin{abstract} 
We study
the decay of scalar fields, in particular the inflaton, into lighter 
scalars in
a De Sitter spacetime background. After providing a practical definition 
of the
rate, we focus on the case of an inflaton interacting with a massless scalar
field either minimally or conformally coupled to the curvature. The evolution
equation for the expectation value of the inflaton is obtained to one loop
order in perturbation theory and the decay rate is recognized from the
solution.  We find the
remarkable result that this decay rate displays an equilibrium 
Bose-enhancement
factor with an effective temperature given by the Hawking temperature 
$H\slash
{2 \pi}$, where $H$ is the Hubble constant. This contribution is 
interpreted as
the ``stimulated emission'' of bosons in a thermal bath at the Hawking
temperature. In the context of new inflation scenarios, we show that inflaton
decay into conformally coupled massless fields slows down the rolling of the
expectation value. Decay into Goldstone
bosons is also studied. Contact with stochastic inflation is
established by deriving the Langevin equation for the coarse-grained
expectation value of the inflaton field to one-loop order in this model. We
find that the noise is gaussian and correlated (colored) and its correlations
are related to the dissipative (``decay'') kernel via a generalized
fluctuation-dissipation relation. 
 
\end{abstract}

\section{Introduction}

Non-equilibrium processes play a fundamental role in cosmological scenarios
motivated by particle physics models. Processes such as thermalization,
reheating and particle decay are important ingredients in most theoretical
scenarios that attempt a description of early universe
cosmology\cite{guth,revs,linde,rocky}.  An important ingredient 
underlying most
attempts at a description of inflationary cosmologies is that of 
equilibration,
which is tacitly assumed whenever quasi-equilibrium methods are invoked in
their study, such as effective potentials and finite temperature field 
theory.

The concept of quasi-equilibrium during an inflationary stage requires a
careful understanding of two different time scales. One is the expansion time
scale $t_H = H^{-1}$, where $H$ is the Hubble parameter, while the other is
the field relaxation time scale, $t_r$. A quasi-equilibrium situation 
obtains whenever
$t_r << t_H$ since in this case particle physics processes occur much 
faster than
the Universe expands allowing fields to respond quickly to any changes in the
thermodynamic variables.

Can the inflaton of inflationary cosmologies be treated as being in
quasi-equilibrium? The answer to this question depends crucially on the
various processes that contribute to the dynamical relaxation of this
field. Some of these processes are the decay of the inflaton into lighter
scalars or fermions, as well as collisions that, while they leave the total
number of particles fixed, will change the phase space distribution of these
particles. 

In this paper we wish to understand the effects of coupling the inflaton to
lighter scalars during the inflationary era. Single particle decay of the
inflaton into lighter scalars or fermions in the post-inflationary era, when
space-time can be well approximated by a Minkowski metric, is by now fairly
well understood within the context of the ``old-reheating''
scenario\cite{dolgov1,chaotic,abbott,kirilova}, but it is much less 
understood
during the {\em early stages} of inflation.

In saying this, we have to distinguish between the various models of
inflation. In ``old'' or ``extended'' inflation\cite{lastein}, the 
inflaton is
trapped in a metastable minimimum, from which it exits eventually via the
nucleation of bubbles of the true vacuum phase. A critical parameter in these
models is the nucleation rate, which is typically calculated assuming 
that the
field is equilibrated\cite{linde,rocky,brand}. Such an assumption will 
only be
justified if the relaxation time of the inflaton in the metastable phase,
i.e. in De Sitter space, is much smaller than the time scale for 
nucleation of
a critical bubble. 

In ``new'' or ``chaotic'' inflationary scenarios\cite{chaotic}, the
supraluminal expansion of the universe is driven by the so-called 
``slow-roll''
dynamics of the inflaton. Typically, this is studied via the zero or finite
temperature effective potential. In doing this, the assumption of
quasi-equilibrium is made, in that one can replace the actual field dynamics,
by that given by a static quantity, the effective potential. This assumption
must be checked, and one way of doing that is by considering the 
evolution of
the inflaton in the presence of couplings to other fields and computing the
field relaxation rate. In particular, if dissipative processes associated 
with
the ``decay'' of the inflaton field into lighter scalars are efficient during
the slow-roll stage in new or chaotic inflationary scenarios, then these may
result in a further slowing down of the inflaton field as it rolls down the
potential hill.

Finally, there is the stochastic inflation approach advocated by
Starobinskii\cite{staro}. In this scenario, inflationary dynamics is studied
with a stochastic evolution equation for the inflaton field with a gaussian
white noise that describes the microscopic fluctuations. As pointed out 
by Hu
and collaborators\cite{lang} and Habib and Kandrup\cite{habib,habib2}, 
noise and
dissipation are related via a fundamental generalized fluctuation-dissipation
theorem and {\em must} be treated on the same footing.

In this article we study the process of relaxation of the inflaton field via
the decay into massless scalar particles in De Sitter space-time with the 
goal
of understanding the modifications in the decay rates and the relevant time
scales as compared to those in Minkowski space-time.  Our study focuses 
on the
description of these processes within ``old/extended '' and ``new'' 
inflationary
scenarios.  Within ``old'' inflation we study the situation in which the
inflaton field oscillates around the metastable minimum (before it eventually
tunnels) and decays into massless scalars.  In the case of ``new'' inflation,
we address the situation of the inflaton rolling down a potential hill, near
the {\em maximum} of the potential, to understand how the process of 
``decay''
introduces dissipative contributions to the inflaton evolution.

We also establish contact with stochastic inflation by {\em deriving} the
Langevin (stochastic) equation for the inflaton field, and analyzing the
generalized fluctuation-dissipation relation between the dissipation kernel
resulting from the ``decay'' into massless scalars and the correlations 
of the
stochastic noise.

At this point we would like to make precise the meaning of ``decay'' in a 
time
dependent background.  The usual concept of the transition rate in Minkowski
space, i.e.  the transition probability per unit volume per unit time for a
particular ({\em in}) state far in the past to an ({\em out}) state far into
the future is no longer applicable in a time dependent background. Relaxing
this definition to finite time intervals ( for example by obtaining the
transition probability from some $t_i$ to some $t_f$) would yield a time
dependent quantity affected by the ambiguities of the definition of particle
states. In Minkowski space-time the existence of time translational 
invariance
allows a spectral representation of two-point functions and the 
transition rate
(or decay rate) is related to the imaginary part of the self-energy on shell.
Such a correspondence is not available in a time dependent background.
Furthermore the time evolution of a scalar field is damped because of the
red-shifting of the energy due to the expanding background even in the 
absence
of interactions. In particular in De Sitter space this damping is exponential
and any other exponential damping arising from interactions will be mixed 
with
the gravitational red-shift.

We propose to {\em define} the ``decay rate'' as the contribution to the
exponential damping of the inflaton evolution due to the interactions with
other fields, which is clearly recognized through the dependence on the
coupling to these fields. This practical definition then argues that in order
to recognize such a ``decay rate'' we must study the real-time dynamics 
of the
non-equilibrium expectation value of the inflaton field in interaction with
other fields, and the identification of this rate should transpire from the
actual evolution of this expectation value.
  
Although the methods presented in this article can be generalized to 
arbitrary
spatially flat FRW cosmologies and include decay of a heavy inflaton into
lighter scalars, we study the simpler but relevant case of inflaton decay 
into
massless scalars in De Sitter space-time. This case affords an analytic
perturbative treatment that illuminates the relevant physical features.

The article is organized as follows: in section II we present a brief
pedagogical review of the closed time path formalism and the derivation 
of the 
non-equilibrium Green's functions for free fields in De Sitter spacetime. In
section III the model is defined and the 
methods to obtain the evolution equation are summarized. Here we study 
several
cases, including minimal and conformal coupling and the slow-roll regime. We
obtain the remarkable result that the decay rate can be understood from
stimulated emission in a bath at the Hawking temperature. In section IV we
study the effect of Goldstone bosons on the inflaton evolution, pointing out
that Goldstone bosons are necessarily always minimally coupled. In 
section V we
make contact with stochastic inflation, derive the stochastic Langevin
equation for the inflaton zero mode and obtain the dissipative kernel and
the 
noise correlation function, which are related by a generalized
fluctuation-dissipation theorem. The resulting noise is gaussian but
colored. In section VI we summarize our results and propose new lines of
study. An appendix is devoted to a detailed derivation, establishing the
consistency of the flat-space limits of the analytic structure of the
dissipative kernel in De Sitter space-time, with Minkowskian calculations.

\section{Non-equilibrium Field Theory}
\subsection{The Closed Time Path Formalism}
 As mentioned in the introduction we will be studying the real-time evolution
of the zero mode of the inflaton in a universe with a time-dependent 
background
metric. Furthermore, we will restrict our investigations to {\em 
perturbative}
phenomena in 
theories in which the inflaton interacts with other matter fields in a De
Sitter universe. We will be primarily concerned with defining and calculating
quantities such as the perturbative ``decay rate'' of the inflaton. For 
studies
of non-perturbative particle production in 
non-equilibrium situations see \cite{boya} (and references
therein). Calculation of particle production in curved spacetime has also 
been
done using the Boguliubov transformations; see for e.g \cite{birrel}

The tools required for studying time-dependent phenomena in quantum field
theories have been available for quite some time, and were first used by
Schwinger {\cite{schwinger}} and Keldysh {\cite{keldysh}}. There are many
articles in the literature using these techniques to study time-dependent
problems {\cite{hu,boya}}. However, these
techniques have not yet become an integral part of the available methods for
studying field theory in extreme environments. We thus present a concise
pedagogical introduction to the subject for the non-practitioner. 

The non-equilibrium
or time dependent description of a system is determined by the time evolution
of the density matrix that describes it. This is in turn described by the
Liouville equation (in the Schr$\ddot{\text{o}}$dinger picture):
\begin{equation}
i{\hbar}\frac{\partial\rho(t)}{\partial t}
=[H(t),\rho(t)].
\end{equation} 
Here we have allowed for an explicitly time dependent Hamiltonian which 
is in
fact the case in an expanding universe. The formal solution of the Liouville
equation is,
\begin{equation}
\rho(t)=U(t)\rho_iU^{-1}(t)
\end{equation}
where $\rho_i$ is the initial density matrix specified at some initial time
$t=0$. The time dependent expectation value of an operator is then easily 
seen
to be,
\begin{equation}
\langle{\cal{O}}\rangle(t)=Tr[\rho_i{U^{-1}(t)\cal{O}}U(t)]\label{expect}
/Tr[\rho_i]
\end{equation}  
In most cases of interest the initial density matrix is either a thermal 
one or
a pure state corresponding to the ground state of some initial 
Hamiltonian. In
either case,
\begin{eqnarray}
&&\rho_i=e^{-\beta H_i}\\
&&H_i=H(t<0).
\end{eqnarray}
The ground state of $H_i$ can be projected out by taking the
$\beta\rightarrow\infty$ limit. To study strongly out of equilibrium
situations it is usually convenient to introduce a time-dependent Hamiltonian
$H(t)$ such that $H(t)=H_i$ for $-\infty\leq t\leq 0$ and $H(t)=H_{evol}(t)$
for $t>0$ where $H_{evol}(t)$ is the evolution hamiltonian that determines
the dynamics of the theory.  

For thermal initial conditions, eq (\ref{expect})
can be written in a more illuminating form by choosing an arbitrary
time $T<0$ for which $U(T)=\exp[-iTH_i]$ so that we may write 
$\exp[-\beta H_i]=\exp[-iH_i(T-i\beta-T)]=U(T-i\beta,T)$. Inserting the
identity $U^{-1}(T)U(T)=1$, commuting $U^{-1}(T)$ with $\rho_i$ and using the
composition property of the evolution operator, we get 
\begin{eqnarray}
\langle{\cal{O}}\rangle(t)&&=Tr[U(T-i\beta,t)
{\cal{O}}U(t,T)]
/Tr[U(T-i\beta,T)]\\\nonumber\\
&&=Tr[U(T-i\beta,T)U(T,T^{\prime})U(T^{\prime},t)
{\cal{O}}U(t,T)]
/Tr[U(T-i\beta,T)]
\end{eqnarray}
where we have introduced an arbitrary large positive time $T^\prime$. The 
numerator now represents
the process of evolving from $T<0$ to $t$, inserting the operator 
${\cal{O}}$,
evolving to a large positive time $T^{\prime}$, and backwards from 
$T^{\prime}$
to $T$ and down the imaginary axis to
$T-i\beta_i$. Eventually one takes 
$T\rightarrow-\infty;T^{\prime}\rightarrow\infty$. This method can be easily
generalised to obtain real time correlation functions of a string of 
operators.

As usual, an operator insertion may be achieved by introducing sources 
into the
time evolution operators, and
taking a functional derivative with respect to the source. Thus we are 
led to
the following generating functional, in terms of time evolution operators in
the presence of sources
\begin{equation}
Z[J^+,J^-,J^\beta]=Tr[U(T-i\beta,T;J^\beta)U(T,T^{\prime};J^-)
U(T^{\prime},T;J^+)]
\end{equation}
with $T\rightarrow -\infty;T^{\prime}\rightarrow\infty$. One can obtain a 
path
integral representation for the generating functional
{\cite{schwinger,keldysh,hu,boya,frwgf,cont}} by inserting a complete
set of field eigenstates between the time evolution operators,
\begin{eqnarray}
Z[J^+,J^-,J^\beta]=&&\int D\Sigma D\Sigma_1 D\Sigma_2
\int{\cal{D}}\Sigma^+{\cal{D}}\Sigma^-{\cal{D}}\Sigma^\beta
e^{i\int^{T^\prime}_T\{{\cal{L}}[\Sigma^+,J^+]-{\cal{L}}[\Sigma^-,J^-]\}}\times
\\\nonumber
&&e^{i\int^{T-i\beta}_T{\cal{L}}[\Sigma^\beta,J^\beta]}.
\end{eqnarray}
The above expression for the generating functional is a path integral defined
on a complex time contour.
Since this path integral actually represents a trace, the field operators at
the boundary points have to satisfy the following conditions:

\begin{eqnarray}
\Sigma^{+}(T)&=&\Sigma^{\beta}(T-i\beta)=\Sigma,\label{ctpboundcond}\\
\Sigma^{+}(T^{\prime})&=&\Sigma^{-}(T^\prime)=\Sigma_2,\nonumber\\
\Sigma^{-}(T)&=&\Sigma^{\beta}(T)=\Sigma_1.\nonumber
\end{eqnarray}
The superscripts $(+)$ and $(-)$ refer to the forward and backward parts 
of the
time contour, while the superscript ($\beta$) refers to the imaginary 
time part
of the contour.
In the limit $T\rightarrow-\infty$ it can be shown
that cross-correlations between fields defined at real times and those 
defined
on the imaginary time contour, vanish due to the $Riemann-Lebesgue$ 
lemma. The
fact that the density matrix is thermal for $t<0$ shows up in the boundary
condition on the fields and the Green's functions. Thus the generating
functional for calculating {\em real time } correlation functions simplifies
to:
\begin{eqnarray}
Z[J^+,J^-]=&&\exp\left\{i\int^{T^\prime}_Tdt[{\cal{L}}_{int}(-i\delta/\delta J^+)
-{\cal{L}}_{int}(i\delta/\delta J^-)]\right\}\times\nonumber\\
&&\exp\left\{\frac{i}{2}\int^{T^\prime}_Tdt_1\int^{T^\prime}_Tdt_2J_a(t_1)J_b(t_2)
G_{ab}(t_1,t_2)\right\}\label{genfunc}
\end{eqnarray}
with $a,b=+,-$.  
We will now proceed to a calculation of the non-equilibrium Green's functions
in De Sitter spacetime
\subsection{Non-equilibrium Green's Functions In De Sitter}

To obtain the non-equilibrium Green's functions we focus on a free scalar 
field
$\Sigma$ in an FRW background metric {\cite{frwgf}}. The generating 
functional
of 
non-equilibrium Green's functions is written in terms of a path integral 
along
the closed time path (CTP) (\ref{genfunc}) with a free Lagrangian
density, ${\cal L}_{0}(\Sigma^{\pm})$ given by
\begin{eqnarray}
{\cal L}_{0}(\Sigma^{\pm}) =\frac{1}{2} [a^{3}(t){\dot{\Sigma}}^{+2}
-a(t){(\vec{\nabla}{\Sigma}^{+})}^2-a^3(t)
(M^2+\xi{\cal R}){{\Sigma}^{+}}^2]
\\ - [+\longrightarrow -]\nonumber
\end{eqnarray}
where ${\cal R} = 6({\ddot{a}}\slash{a}+{{\dot{a}^2}}\slash{a^2})$. 

We prepare the system such that it is thermal for $t<0$ with temperature
$1\slash 
\beta$. At $t=0$, we switch the interactions on and follow the resulting time
evolution of the coupled fields. As seen in the previous section, the CTP
formulation of non-equilibrium field theory imposes certain boundary 
conditions
on the 
fields (\ref{ctpboundcond}).

Rewriting the Lagrangian as a quadratic form in $\Sigma^{\pm}$ 
we obtain the Green's function equation,
\begin{equation}
[\frac{\partial^2}{\partial t^2}+3\frac{\dot{a}}{a}\frac{\partial}{\partial
t}-\frac{{\nabla}^2}{a^2} +(M^2+\xi{\cal R})]G(x,t;x^{\prime},t^{\prime}) =
\frac{\delta^4({x}-{x}^{\prime})}{a^{3/2}(t)a^{3/2}(t^{\prime})}.
\end{equation}

We use spatial translational invariance to define,
\begin{equation}
G_{k}(t,t^{\prime}) = {\int}d^{3}xe^{i\vec{k}\cdot 
\vec{x}}G(x,t;0,t^{\prime})
,   
\end{equation}
so that the spatial Fourier transform of the Green's function obeys,
\begin{equation}
[\frac{d^2}{dt^2}+3\frac{\dot{a}}{a}\frac{d}{dt}+
\frac{k^2}{a^2}+(M^2+\xi{\cal R})]G_{k}(t,t^{\prime}) 
= \frac{\delta(t-t^{\prime})}{a^{3/2}(t)a^{3/2}(t^{\prime})} .
\end{equation}
The solution can be cast in a more familiar form by writing
\begin{equation}
G_{k}(t,t^{\prime})=\frac{f_{k}(t,t^{\prime})}
{a^{3/2}(t)a^{3/2}(t^{\prime})},\label{gfscale}
\end{equation} 
so that the function $f_k(t,t')$ obeys the second order differential 
equation,
\begin{equation}
[\frac{d^2}{dt^2}-(\frac{3{\ddot{a}}}{2a}+
\frac{3{\dot{a}}^2}{4a^2})+\frac{k^2}{a^2}+
(M^2+\xi{\cal R})]f_{k}(t,t^{\prime})=\delta(t-t^{\prime}) .
\end{equation}
The general solution of this equation is of the form
\begin{equation}
f_{k}(t,t^{\prime})={f_{k}}^{>}(t,t^{\prime})\Theta(t-t^{\prime})
+{f_{k}}^{<}(t,t^{\prime})\Theta(t^{\prime}-t) ,
\end{equation}
where ${f_{k}}^{>}$ and ${f_{k}}^{<}$ are solutions to the homogeneous 
equation
obeying the appropriate boundary conditions. They can be expanded in 
terms of
normal mode solutions to the homogeneous equation:
\begin{eqnarray}
{f_{k}}^{>}(t,t^{\prime}) = A^{>}U_{k}(t)U^{\ast}_{k}(t^{\prime})
+B^{>}U_{k}(t^{\prime})U^{\ast}_{k}(t) ,\label{definef}
\\{f_{k}}^{<}(t,t^{\prime}) = A^{<}U_{k}(t)U^{\ast}_{k}(t^{\prime})+
B^{<}U_{k}(t^{\prime})U^{\ast}_{k}(t) .\nonumber
\end{eqnarray}
The choice of mode functions can be made unique for a given ${\xi}$ and 
$M$ by
requiring that,
\begin{eqnarray}
&&\lim_{k\rightarrow\infty} U_{k}(t) \sim e^{-ikt} ,\label{eqlim}\\
&&\text{or},\nonumber\\
&&\lim_{H\rightarrow0} U_{k}(t)
\sim e^{-i\omega_{k}t} ,\nonumber
\\\nonumber\\
&&\omega_{k}=\sqrt{k^2+M^2} .\nonumber
\end{eqnarray}
Here $H\equiv \dot{a(t)}\slash {a(t)}$ is the Hubble parameter. Stated 
simply,
these conditions require that we must necessarily recover the rules of field
theory in Minkowski space when we take the flat-space or the short distance
($k\rightarrow\infty$) limits. These boundary conditions are similar to those
invoked by Bunch and Davies\cite{bunch,birrel}. For the De Sitter space case,
with $a(t)=\exp(H t)$, the mode functions are the Bessel
functions\cite{birrel}, $J_{\pm\nu}(ke^{-Ht}/H)$ provided $\nu$ is not an
integer, where
\begin{equation}
\nu = i{\sqrt{\frac{M^2}{H^2}+12{\xi}-{\frac{9}{4}}}} \label{eqnu} . 
\end{equation}
For the values of ${\xi}$ and $M$ for which ${\nu}$ becomes real these Bessel
functions become manifestly real and the solutions that fulfill the boundary
conditions (\ref{eqlim}) are obtained as linear combinations of these
functions. In order to obtain the full Green's functions the constants
$A^{>},B^{>},A^{<},B^{<}$ (\ref{definef}) are determined by implementing the
following 
boundary conditions\cite{hu,boya,frwgf} :
\begin{eqnarray}
&&f^{>}_{k}(t,t)=f^{<}_{k}(t,t)\label{bc}                                     
\\&&{\dot{f}}^{>}_{k}(t,t)-{\dot{f}}^{<}_{k}(t,t)=1\nonumber
\\&&f^{>}_{k}(T-i\beta,t)=f^{<}_{k}(T,t)\nonumber
\end{eqnarray}
The first two conditions are obtained from the continuity of the Green's
function and the jump discontinuity in its first derivative. The third 
boundary
condition is just a consequence of the periodicity of the fields in imaginary
time, which followed from the assumption that the density matrix is that 
of a
system in thermal equilibrium at time $T<0$. Although we allowed for a 
thermal
initial density matrix, we will focus only on the zero temperature case, 
since
during the De Sitter era the temperature is rapidly red-shifted to zero. The
most general case for the Green's function can be found in\cite{frwgf}.
Implementing these conditions for the case of imaginary $\nu$ and zero
temperature ($\beta\rightarrow\infty$) we find :

\begin{eqnarray}
G^{>}_{k}(t,t^{\prime})&=&\frac{-\pi J_{\nu}(z)J_{-\nu}(z^{\prime}) 
}{2H{\sin}(\pi\nu)
e^{3Ht/2}e^{3Ht^{\prime}/2}}
 \label{gregreat1} \\
G^{<}_{k}(t,t^{\prime})&=&\frac{-\pi J_{-\nu}(z)
J_{\nu}(z^{\prime})}{2H{\sin}(\pi\nu)e^{3Ht/2}
e^{3Ht^{\prime}/2}},\nonumber
\\ z&=&k\frac{e^{-Ht}}{H},\hspace{.4in}z^{\prime}=
k\frac{e^{-Ht^{\prime}}}{H}\label{greless1}
\end{eqnarray}
and $\nu$ is given by (\ref{eqnu}). 

In the case of interest for this article, that of a massless scalar field
either conformally or minimally coupled to the curvature, the Bessel 
functions
in the above expressions must be replaced by the corresponding Hankel
functions.These two cases correspond to:

\begin{eqnarray}
&& (i)M=0, {\xi}=0,\text{ }\Rightarrow \text{ }U_{k}(t)
= {H^{(1)}}_{3/2}(ke^{-Ht}/H), \label{minimalcoup} \\
&& (ii)M=0, {\xi}=\frac{1}{6},\text{ }\Rightarrow 
\text{ }U_{k}(t)= {H^{(1)}}_{1/2}(ke^{-Ht}/H), \label{conformalcoup}
\end{eqnarray}
Finally, the Green's functions for these
two cases of a massless scalar field, in terms of $z$ and $z^{\prime}$, are:
\begin{eqnarray}
&& G^{>}_{k}(t,t^{\prime}) = \frac{-i\pi H^{(1)}_{\nu}(z)H^{(2)}_{\nu} 
(z^{\prime})}{4He^{3Ht/2}e^{3Ht^{\prime}/2}} \label{ggreater} \\
&& G^{<}_{k}(t,t^{\prime}) = \frac{-i\pi H^{(2)}_{\nu}(z)H^{(1)}_{\nu}
(z^{\prime}) }{4He^{3Ht/2}e^{3Ht^{\prime}/2}}
\label{glesser}
\end{eqnarray}
with $\nu=3/2$ for minimally coupled or $\nu=1/2$ for conformally 
coupled. From
the expressions for $G^{>}_{k}$ and $G^{<}_{k}$ we can compute the 
required 2-point
functions:
\begin{eqnarray}
i&&{\langle}T\Sigma^{+}(\vec{r},t)\Sigma^{+}({\vec{r}}^{ \prime},t^{\prime})
{\rangle}=G^{>}(\vec{r},t;{\vec{r}}^{ \prime},t^{\prime})
\Theta(t-t^{\prime})+G^{<}(\vec{r},t;{\vec{r}}^{ \prime},t^{\prime})
\Theta(t^{\prime}-t) ,\label{gplpl}\\
i&&{\langle}T\Sigma^{-}(\vec{r},t)\Sigma^{-}
({\vec{r}}^{ \prime},t^{\prime}){\rangle}
=G^{>}(\vec{r},t;{\vec{r}}^{ \prime},t^{\prime})
\Theta(t^{\prime}-t)+G^{<}(\vec{r},t;{\vec{r}}^{ \prime},t^{\prime})
\Theta(t-t^{\prime}) ,
\label{glele}\\
-i&&{\langle}T\Sigma^{+(-)}(\vec{r},t)\Sigma^{-(+)}
({\vec{r}}^{ \prime},t^{\prime}){\rangle}=G^{<(>)}(\vec{r},t;{\vec{r}}
^{\prime},t^{\prime}) .
\label{gplmin}
\end{eqnarray}

In our subsequent analysis we will always work in the 
$\beta\rightarrow\infty$
limit (zero initial temperature) unless stated otherwise. 

\section{the model and the method}
We now turn to the main topic of this work, the study of an inflaton,
described by a field $\Phi$ coupled to a massless scalar field $\sigma$ with
the non-equilibrium Lagrangian density

\begin{eqnarray}
 {\cal L} =\frac{1}{2} [e^{3Ht}{\dot{\Phi}}^{+2}-e^{Ht}
{(\vec{\nabla}{\Phi}^{+})}^2-e^{3Ht}
(m_{\Phi}^2+12{\xi}_{\Phi}H^2){\Phi}^{+2}\\\nonumber
+e^{3Ht}{\dot{\sigma}}^{+2}-e^{Ht}
{(\vec{\nabla}{\sigma}^{+})}^2-12e^{3Ht}{\xi}_
{\sigma}H^2{\sigma}^{+2}]+\frac{1}{2}{e^{3Ht}}
[g{\Phi}^{+}{\sigma}^{+2}+h(t){\Phi}^{+}]\nonumber
\\
-[+\longrightarrow-].\label{lagrangian}
\end{eqnarray}

We have introduced a ``magnetic'' (source) term $h(t)$ for two reasons:

\begin{enumerate}
\item{such
a term will be generated by renormalization from tadpole graphs, 
therefore we
introduce $h(t)$ as a counterterm to cancel these contributions,}

\item{$h(t)$ will serve as a
Lagrange multiplier to define our problem as an initial condition problem.
That is, we will fix $h(t)$ such that the divergences from tadpole graphs
contributing an 
inhomogeneous term to the equation of motion are cancelled and the initial
value of the inflaton field is fixed for $t<0$. Such a procedure will be 
implemented
below when we obtain the equation of motion for the expectation value of the
inflaton.}
\end{enumerate}

Clearly, in De Sitter space-time the coupling to the curvature only 
serves to
redefine the mass of the fields. For the inflaton, we just absorb this 
term in
a redefinition of the mass, and set ${\xi}_{\Phi}=0$ without any loss of
generality.

To study the time evolution of the expectation value of the inflaton 
field we
invoke the tadpole method(see Ref.\cite{boya} and references therein) to 
obtain
its equation of motion.  This is implemented by first shifting ${\Phi}$ 
by its
expectation value in the non-equilibrium state as follows :

\begin{equation}
\Phi^{\pm}(\vec{x},t)=\phi(t)+\psi^{\pm}(\vec{x},t) \; \; \; \; \\;\; \; \;
\phi(t)=\langle\Phi^{\pm}(\vec{x},t)\rangle\label{shift}
\end{equation}
and then enforcing the tadpole condition which is simply a consequence of
(\ref{shift})

\begin{equation}
\langle\psi^{\pm}(\vec{x},t)\rangle=0.\label{tadpolecond}
\end{equation}
$\phi(t)$ being the spatially homogeneous inflaton zero mode that drives
inflation. We also require that
$\langle\sigma^{\pm}(\vec{x},t)\rangle=0$ to all orders, which means that the
$\sigma$ field does not acquire an expectation value.  After the shift of the
inflaton field (\ref{shift}), the non-equilibrium action reads:

\begin{eqnarray}
S=&&\int d^3xdt\left\{{\cal{L}}_0(\psi^+)+{\cal{L}}_0(\sigma^+)+
\psi^+e^{3Ht}\left[-\ddot{\phi}-3H\dot{\phi}-m^2_{\Phi}\phi\right]
+e^{3Ht}\frac{g}{2}\phi(\sigma^+)^2\right.\label{noneqaction}\\\nonumber
&&\left.+e^{3Ht}\frac{g}{2}\psi^+(\sigma^+)^2+e^{3Ht}\frac{h}{2}\psi^+
-\left(+\rightarrow-\right)\right\}
\end{eqnarray}
where ${\cal{L}}_0$ represents the free theories for the $\psi$ and $\sigma$
fields. The tadpole condition (\ref{tadpolecond}) is implemented by 
calculating
the expectation value of the field operator $\psi^{\pm}$ by 
inserting it into the non-equilibrium path integral and expanding in 
powers of 
$g$, about the free theory and setting the resulting expression to zero. 
In particular we evaluate the following expression to order $g^2$ in
perturbation theory:

\begin{eqnarray}
&&\langle\psi^+(\vec{x},t)\rangle=\int {\cal{D}}[\psi^\pm]{\cal{D}}
[\sigma^\pm] 
e^{i(S_0[+]-S_0[-])}\times\psi^+(\vec{x},t)\times\\\nonumber
&&\exp\left\{\int d^3x^\prime dt^\prime
\left[e^{3Ht^\prime}\left[-\ddot{\phi}-3H\dot{\phi}- 
m^2_{\Phi}\phi\right]\psi^+
+e^{3Ht^\prime}\frac{g}{2}\phi(\sigma^+)^2
+e^{3Ht^\prime}\frac{g}{2}\psi^+(\sigma^+)^2+e^{3Ht^\prime}\frac{h}{2}\psi^+
\right.\right.\\\nonumber
&&\left.\left.\hspace{5 in}-\left(+\rightarrow-\right)\right]\right\}
\\\nonumber
&&\hspace{.7 in}=0
\end{eqnarray}
The equation of motion is obtained by independently setting the 
coefficients of
$\langle \psi^+(x)\psi^+(x^{\prime})\rangle$ and 
$\langle\psi^+(x)\psi^-(x^{\prime})\rangle$ to zero. This is justified 
because
these two correlators are independent functions. It is easily seen that the
condition $\langle\psi^-\rangle=0$ yields exactly the same equations.

This
procedure gives rise to an equation of motion for the zero mode $\phi(t)$ and
is in fact equivalent to extremizing the one-loop effective action which
incorporates the equilibrium boundary conditions at $t<0$.

It is worth noting that this scheme, referred to as {\em amplitude 
expansion}, 
also assumes that the amplitude of the zero-mode $\phi(t)$ is
``small'' so that terms such as $g\phi(t)\sigma^2$ may be treated
perturbatively.   

To one-loop order (${\cal{O}}(g^2)$) we obtain the following form of the
equation of motion.

\begin{eqnarray}
&& 
\ddot{\phi}+3H\dot{\phi}+m_{\Phi}^2\phi-\frac{g}{2}\langle\sigma^2\rangle(t)
-i\frac{g^2}{4{\pi}^2}\int^{t}_{-\infty}d{\tau}K(t,\tau)\phi(\tau)- 
\frac{h(t)}{2}=0,\\
&& K(t,\tau)={\int}^{\infty}_{0}{k^2}dk
[G^{>2}_{k}(\tau,t)-G^{<2}_{k}(\tau,t)]e^{3H\tau}\label{kernel},
\end{eqnarray}
Spatial translational invariance guarantees that $\langle 
\sigma^2\rangle$ is
only time dependent.
We can combine the $\phi$ independent
contributions to the equation of motion into a single function:

\begin{equation}
\tilde{h}(t)=\frac{h(t)}{2}+
\frac{g}{2}\langle \sigma^2\rangle(t).
\end{equation}
If the $\sigma$ is conformally coupled to the curvature $\langle
\sigma^2\rangle$ represents a quadratically divergent piece which must be
subtracted away by a renormalization of the ``magnetic field'' $h$.
On the other hand if $\sigma$ is minimally coupled, one obtains the well 
known
linearly growing contribution \cite{linde} after subtracting away the
divergence leading to 
a modified inhomogeneous term in the zero mode equation of motion: 
\begin{equation}
\tilde{h}(t)= \frac{g H^3 t}{8\pi^2} +\frac{h(t)}{2}\label{tadp}.
\end{equation}
We will, however, show explicitly in the next section that this term does 
{\em not}
enter in the calculation of the perturbative ``decay rate'' in this theory.

We use
$\tilde{h}(t)$ as a Lagrange multiplier to enforce the constraint that for
$t<0,\ \phi(t)=\phi_i,\ \dot{\phi}(t)=0$ is a solution of the effective
 equation of motion
for the expectation value. In this manner, we define an initial
value problem  with Cauchy data on a space-like surface 
for the dynamics in which the expectation value of the inflaton
field is ``released'' from some initial value at time $t=0$. For $t>0$
the equation of motion becomes

\begin{eqnarray}
&& \ddot{\phi}(t)+3H\dot{\phi}(t)+m_{\Phi}^2 \phi(t)
-i\frac{g^2}{4{\pi}^2}\int^{t}_{0}d{\tau}K(t,\tau)\phi(\tau)=
 \tilde{h}(t) 
\label{eqnofmotion} \\
&& \phi(t=0)=\phi_i \; , \; \dot{\phi}(t=0) = 0 \label{iniconds}
\end{eqnarray}
with the kernel given by (\ref{kernel}). 


We now proceed to study the solution
to this equation of motion in the cases of conformal and minimal coupling.
 
\subsection{Conformally coupled massless $\sigma$:}

For a conformally coupled massless scalar, the Green's functions are 
given by
(\ref{ggreater},\ref{glesser}) with the Hankel functions as in
(\ref{conformalcoup}).

The kernel has a particularly simple form in this case:

\begin{equation}
K(t,\tau)=-i\int^{\infty}_{0}dk
\frac{\sin[2k(e^{-H\tau}-e^{-Ht})/H]}{2e^{2Ht}e^{-H\tau}}.
\end{equation}
The $k$-integral can be done by introducing a small convergence factor as
follows :

\begin{eqnarray}
\int^{\infty}_{0}\sin({\alpha}k)dk & = & \lim_{\epsilon \rightarrow 0}
Im\left[\int^{\infty}_{0}e^{i(\alpha+i\epsilon)k}dk\right] \nonumber \\
&=& \lim_{\epsilon \rightarrow 0}\frac{\alpha}{\alpha^2+\epsilon^2}.
\end{eqnarray}
Implementing the above prescription and absorbing a positive finite piece 
into
$\epsilon$ yields,

\begin{eqnarray}
K(t,\tau)=-i\frac{H}{4}\frac{[e^{H(t-\tau)}-1]}
{(e^{H(t-\tau)})[(e^{H(t-\tau)}-1)^2+\epsilon^2]} \label{finalkernel}.
\end{eqnarray}
The equation of motion then becomes,
\begin{equation}
\ddot{\phi}+3H\dot{\phi}+m_{\Phi}^2\phi-
\frac{g^2H}{16{\pi}^2}\int^{t}_{0}d{\tau}
\frac{[e^{H(t-\tau)}-1]}{(e^{H(t-\tau)})
[(e^{H(t-\tau)}-1)^2+\epsilon^2]}\phi(\tau)=\tilde{h}(t).
\label{fineqofmotion}
\end{equation}

A remarkable point to note, is that although time translation is {\em 
not} a
symmetry in De Sitter space-time, nonetheless the term inside the 
integral is
in 
the form of a convolution, and the equation, being linear in $\phi$, can be
solved via Laplace transforms. This is presumably due to the $O(4,1)$ 
symmetry
of De Sitter space. 

Notice that $\epsilon$ serves as a short-distance regulator, and the
$\epsilon\rightarrow 0$ limit cannot be taken inside the integral because 
doing
so results in a logarithmic short distance singularity in the equation of
motion. This divergence is isolated by performing an integration by parts,
keeping the $\epsilon$ dependence, resulting in the equation of motion

\begin{eqnarray}
&&\ddot{\phi}(t)+3H\dot{\phi}(t)+m_{\Phi}^2\phi(t)-\frac{g^2H}{16{\pi}^2}
\left\{-\frac{\phi (t)}{H}\ln\epsilon-\frac{\phi(t)}{H}+\frac{\phi_i}{H}
\left[\ln(1-e^{-Ht})+e^{-Ht}\right] \right.  \nonumber \\
&& \left. +{\int}d{\tau}\frac{\dot{\phi}(\tau)}{H} \xi(t-\tau) \right\}=
\tilde{h}(t)  \label{diveqn} \\
&& \xi(t-\tau) = \left[\ln(1-e^{-H(t-\tau)})+e^{-H(t-\tau)}\right].
 \label{sigmakernel}
\end{eqnarray}
This expression makes it clear that the divergence can be absorbed in a mass
renormalization,

\begin{equation}
m^{2}_{\Phi,R}=m^{2}_{\Phi}+
\frac{g^2}{16{\pi}^2}(\ln\epsilon+1). \label{renormass}
\end{equation}
In what follows we will refer to $m_{\Phi}$ as the renormalized mass to 
avoid 
cluttering the notation.

Taking the Laplace transform of the equations of motion, defining

\begin{eqnarray}
\tilde{\phi}(s)      & = & \int^{\infty}_{0}e^{-st}\phi(t)dt
 \label{filapla} \\
\tilde{\xi}(s) & = &  \int^{\infty}_{0}e^{-st}\xi(t)dt \label{sigmalapla}\\
\tilde{h}(s) & = & \int^{\infty}_{0}e^{-st}\tilde{h}(t)dt \label{hlapla}
\end{eqnarray}
and using the initial condition $\dot{\phi}(t=0)=0$, we find

\begin{eqnarray}
&& s^2\tilde{\phi}(s)-s\phi_i +3H(s\tilde{\phi}(s)-\phi_i)
+m^{2}_{\Phi,R}\tilde{\phi}(s)+\frac{g^2}{16{\pi}^2}{\tilde{\Sigma}}(s)\tilde{\phi}(s)
=\tilde{h}(s) \label{lapla}\\
&& \tilde{\Sigma}(s) = -s\int^{\infty}_{0}dte^{-st}[\ln(1-e^{-Ht})+e^{-Ht}]
=-s\tilde{\xi}(s)  
\label{laplakernel}
\end{eqnarray}

\subsection{The Gibbons-Hawking Temperature}

It is well known that fields in De Sitter space can be thought of, in some
contexts, as being embedded in a thermal bath at a temperature $H\slash {2
\pi}$, the so-called Gibbons-Hawking temperature\cite{gibhawk}. How do we see
these effects in the context of our calculation?

First we notice that the Green's functions(\ref{gregreat1},\ref{greless1},
\ref{ggreater},\ref{glesser}) and the kernels (\ref{finalkernel}, 
\ref{sigmakernel}) 
in (\ref{diveqn}) obey the periodicity condition in 
{\em imaginary time}
\begin{equation}
K(t-\tau)=K(t-\tau+\frac{2\pi i}{H}). 
\end{equation}
These kernels can be
analytically continued to imaginary time $t-\tau=-i\eta$ and expanded in a
discrete series in terms of the Matsubara frequencies
\begin{equation}
\omega_n  =  \frac{2\pi n}{\beta_H} \; , \; \;\frac{1}{\beta_H}= T_H  = 
 \frac{H}{2\pi} \label{matsufreq}
\end{equation}
$T_H$ is recognized as the Gibbons-Hawking temperature in this space-time
\cite{birrel}. Then as a function of imaginary time (\ref{sigmakernel}) 
can be
written as
\begin{eqnarray}
\xi(\eta)     & = & \frac{1}{\beta_H} \sum_{n=-\infty}^{\infty}
e^{i\omega_n \eta} \Xi(\omega_n)  
\label{finitetexp} \\
\Xi(\omega_n) & = & \int_{0}^{\beta_H}d\eta e^{-i\omega_n \eta}\xi(\eta) 
\label{xiofomega}
\end{eqnarray}
We find 
\begin{equation}
\Xi(\omega_n) = -\beta_H \left(\frac{H}{\omega_n}-\delta_{n,1}\right),\ n
\geq 1.
\end{equation}

>From this expression, $\tilde{\Sigma}(s)$ can be obtained at once:
\begin{equation}
 \tilde{\Sigma}(s) =-s\tilde{\xi}(s)=
\sum^{\infty}_{1}H(\frac{1}{nH}-\frac{1}{s+nH})-\frac{s}
{s+H}.\label{laplafin}
\end{equation} 
This Laplace transform has simple poles at minus the Matsubara 
frequencies 
corresponding to the Hawking temperature.

Finally  we obtain the solution for the Laplace transform
\begin{equation}
\tilde{\phi}(s)=\frac{\phi_i (s+3H)+\tilde{h}(s)}{s^2+3Hs+m^{2}_{\Phi}+
\frac{g^2}{16{\pi}^2}{\tilde{\Sigma}}(s)}. \label{laplasoluti}
\end{equation} 
This expression clearly shows that the information on the ``decay rate''
is obtained from the imaginary part of the poles of the denominator and
that the ``effective magnetic field'' $\tilde{h}(t)$ whose Laplace
transform $\tilde{h}(s)$ is explicit in eq.(\ref{laplasoluti}) is not
relevant to understand this ``decay rate''. In fact as we will show shortly,
the ``decay rate'' is determined only by the ``self-energy'' $\Sigma(s)$. 
Since
our goal is to understand 
the ``decay rate'' we set $\tilde{h}(s)=0$ without loss of generality. 
 
The denominator in the above expression is the inverse propagator for the
massive scalar, evaluated at zero spatial momentum and 
$\tilde{\Sigma}(s)$ is
the one-loop correction to the self-energy. However, from 
(\ref{laplafin}) we
see that $\tilde{\Sigma}(s)$ is analytic in the $s$-plane with simple 
poles at
$s=-nH$, with $n$ an integer$\neq 1$). This is a surprising result, 
because 
we expect that,
as in Minkowski space, there would be a cut in the $s$-plane, which since
the particles in the loop are massless, would run along the entire imaginary
axis.  Obviously, this is not what happens - the analytic structure is {\em
very different} in De Sitter space-time from that in Minkowski 
space-time. In
fact we find that $\tilde{\phi}(s)$ has two simple poles, perturbed from 
their
original positions by ${\cal O}(g^2)$ corrections. We will understand this
result by taking the flat-space limit in a later section.

The real time dynamics of the zero mode of the inflaton is now found by
inverting the Laplace transform by complex integration:
\begin{equation}
\phi(t)=\int^{c+i\infty}_{c-i\infty}e^{st}\tilde{\phi}(s)\frac{ds}{2{\pi}i}
\end{equation}
where the Bromwich contour is taken to the right of all the 
singularities. 

The poles of $\tilde{\phi}(s)$ can be obtained by setting,
\begin{equation}
s^2+3Hs+m^{2}_{\Phi}+\frac{g^2}{16{\pi}^2}\tilde{\Sigma}(s)=0. \label{poles}
\end{equation}

We analyze the $4m^{2}_{\Phi}>9H^2$ and $4m^{2}_{\Phi}<9H^2$ cases
separately. In the first case, with $4m^{2}_{\Phi}>9H^2$, we have that in the
absence of interactions the inflaton undergoes damped oscillations 
red-shifted
by the expansion. At zeroth order the poles (\ref{poles}) are at

\begin{equation}
s^{\pm}_{0}=\frac{-3H}{2}{\pm}i\frac{\sqrt{4m^{2}_{\Phi}-9H^2}}{2}.
\end{equation}
We absorb the real part of the self energy {\em on shell} in a further 
(finite)
redefinition of the renormalized mass. Since $\tilde{\Sigma}(s)$ is a
meromorphic function of s, $Re[\tilde{\Sigma}(s^{+}_{0})]=
Re[\tilde{\Sigma}(s^{-}_{0})]$. Then to this order the ``pole mass'' is given
by

\begin{equation}
m^{*2}_{\Phi}=
m^{2}_{\Phi}+\frac{g^2{Re[\Sigma}(s^{\pm}_{0})]}{16{\pi}^2}.
\end{equation}
Obviously this mass is independent of the renormalization scheme. We also
define the subtracted self energy
$\tilde{\Sigma}^{*}(s)=\tilde{\Sigma}(s)-Re[\tilde{\Sigma}(s^{+}_{0})]$, 
and at
this stage it is convenient to introduce the ``effective mass''

\begin{equation}
M^2_{\Phi} = m^{*2}_{\Phi}-\frac{9H^2}{4}. \label{effmass}
\end{equation}
in terms of which, to order ${g^2}$, we find

\begin{eqnarray}
s^{\pm}&=&\frac{-3H}{2}{\pm}i \sqrt{M^{2}_{\Phi}
+\frac{g^2}{16{\pi}^2}\tilde{\Sigma}^{*}(s^{+})}\nonumber \\
&=&\frac{-3H}{2}{\pm}i M_{\Phi}
{\pm}i\frac{g^2 Im\tilde{\Sigma}(s^{\pm}_{0})}{32{\pi}^2 M_{\Phi}}
+{\cal O}(g^4)
\end{eqnarray}

The Bromwich countour for the inverse Laplace transform can now be 
deformed by
wrapping around the imaginary axis and picking up the poles\cite{boya}. The
resulting expression has the Breit-Wigner form of a sharp resonance 
centered at
the ``pole mass'' that is very narrow in the weak coupling limit, leading to
the time evolution of the inflaton given by

\begin{equation}
\phi(t)=Z\phi_i e^{-3Ht/2}e^{-{\Gamma}t/2}\cos(M_{\Phi}t+\alpha)
\end{equation}
where,
\begin{eqnarray}
&&Z^2=1+\frac{9H^2}{4M^2_{\Phi}}-\frac{2g^2B^{\prime}}{32{\pi}^2M_{\Phi}}-
\frac{12g^2HB}{128{\pi}^2M^3_{\Phi}}-\frac{18g^2H^2B^{\prime}}{128{\pi}^2M^3_
{\Phi}}\\\nonumber
\\\nonumber   
&&\alpha=\arctan\left[-\frac{3H}{2M_{\Phi}}+\frac{g^2}{16{\pi}^2}
[\frac{A^{\prime}}{2M_{\Phi}}+\frac{9H^2A^{\prime}}{8M^3_{\Phi}}-
\frac{18H^2B}{16M^4_{\Phi}}]\right]
\\\nonumber
\\\nonumber
&&B=Im\left[\tilde{\Sigma}^{*}(s)\right]_{s=\frac{-3H}{2}+iM_{\Phi}}\\\nonumber
\\\nonumber
&&A^{\prime}=Re\left[\frac{\partial\tilde{\Sigma}^{*}(s)}
{{\partial}s}\right]_{s=\frac{-3H}{2}+iM_{\Phi}}
\\\nonumber
\\\nonumber
&&B^{\prime}=Im\left[\frac{\partial\tilde{\Sigma}^{*}(s)}
{{\partial}s}\right]_{s=\frac{-3H}{2}+iM_{\Phi}}.
\\\nonumber
\end{eqnarray}
The most interesting of these quantities is the damping rate, given
by
\begin{equation}
\Gamma=\frac{g^2Im\tilde{\Sigma}^{*}(s^{+}_{0})}{16{\pi}^2 M_{\Phi}}\label
{dampin}. 
\end{equation}
The imaginary part of the self-energy in the above expression, can be 
evaluated
using standard sum formulas (these sums also appear finite temperature field
theory\cite{dolan,kapusta}) and is found to be

\begin{equation}
\Gamma=
\frac{g^2\tanh\left(\frac{\beta_H\sqrt{m^{*2}_{\Phi}-\frac{9H^2}{4}}}{2}
\right)}{32\pi\sqrt{m^{*2}_{\Phi}-\frac{9H^2}{4}}}
=\frac{g^2\tanh\left(\frac{\beta_H M_{\Phi}}{2}\right)}{32\pi M_{\Phi}}
 \label{rate1}
\end{equation}
This expression for $\Gamma$ is a monotonically increasing function of
$H$. When $H=0$, it matches with the flat-space decay rate
\begin{equation}
\Gamma_{Minkowski}=\frac{g^2}{32{\pi}m_{\Phi}}.
\end{equation}
We thus obtain the result that the inflaton decay proceeds more
rapidly in a De Sitter background than in Minkowski space-time.

The rate (\ref{rate1}) can be written in a more illuminating manner as

\begin{eqnarray}
&& \Gamma=\frac{g^2\coth\left(\frac{\beta_H \omega_0}{2}\right)}{32\pi 
M_{\Phi
}}=
\frac{g^2\left[1+2 n_b(\omega_0)\right]}{32\pi M_{\Phi}} \label{bose1} \\
&&\omega_0= -is^{+}_0 \; \; ; \; \; n_b(\omega_0) = \frac{1}{e^{\beta_H 
\omega
_0}-1}
\label{bosefactor}
\end{eqnarray}
This is a remarkable result - the decay rate is almost the same as that in
Minkowski space, but in a thermal bath at the Hawking
temperature\cite{boya,note}. Habib\cite{habib2} has also found an intriguing
relationship 
with the Hawking temperature in the probability distribution functional 
in his
studies of  
stochastic inflation. 

If we now look at the $4m^{2}_{\Phi}<9H^2$ case, we see that the inflaton
ceases to propagate, i.e. its Compton wavelength approaches the horizon size
and there is no oscillatory behavior in the classical evolution of the zero
mode:
\begin{eqnarray}
\phi(t)&=&(\frac{9H^2}{4m^{2}_{\Phi}}-1)^{-1/2}\phi_ie^{-3Ht/2}
\sinh\left(|M_{\Phi}|t+\beta\right)
\label{clas2} \\
\tanh \beta&=&\frac{2|M_{\Phi}|}{3H}.\nonumber
\end{eqnarray}
The one loop contribution is the same as in the previous case  but now  
the 
poles of $\tilde{\phi}(s)$
 lie on the real axis:
\begin{eqnarray}
s^{\pm}_{0}&=&\frac{-3H}{2}{\pm}\frac{\sqrt{9H^2-4m^{2}_{\Phi}}}{2}.\\\nonumber
s^{\pm}&=&\frac{-3H}{2}{\pm}\frac{\sqrt{9H^2-4m^{2}_{\Phi}-
4\frac{g^2}{16{\pi}^2}{\tilde{\Sigma}}(s^{\pm})}}{2}\\\nonumber
\end{eqnarray}
We define 
\begin{equation}
\Sigma(s^{\pm}_{0})=C{\pm}D
\end{equation}
and absorb C into a finite mass renormalization,
 \begin{equation} 
m^{*2}_{\Phi}=
m^{2}_{\Phi}+\frac{g^2{C}}{16{\pi}^2}.
\end{equation}
Inverting the transform we 
obtain,                                                    
\begin{equation}
\phi(t)= Z\phi_ie^{-3Ht/2}e^{-{\Gamma}t/2}\sinh(\frac{K}{2}t
+\beta),
\end{equation}
with $Z$ being the (finite) wave function renormalization i.e. the 
residue at the poles, and
\begin{eqnarray}
K&=&\sqrt{9H^2-4m^{*2}_{\Phi}}\label{kprime} \\
\Gamma&=&\frac{g^2\tan(\frac{\pi}{2}\sqrt{9-
\frac{4m^{*2}_{\Phi}}{H^2}})}{16{\pi}\sqrt{9H^2-4m^{*2}_{\Phi}}}
\hspace{0.4in}\text{for}\hspace{0.4in}
\frac{m^{*}_{\Phi}}{\sqrt{2}}>H>\frac{2m^{*}_{\Phi}}{3}.\label{newgamma}
\end{eqnarray}
We also see that the numerator of the damping rate in eq.(3.46) diverges at
$H=\frac{m^{*}_{\Phi}}{\sqrt{2}}>\frac{2m^{*}_{\Phi}}{3}$, which indicates
the breakdown of perturbation theory.

The decay rate may again  be written in the a form that makes explicit 
the 
effect of the
Hawking temperature and the ``stimulated decay'' with the Bose-Einstein 
distribution function at the Hawking temperature.
\begin{equation}
\Gamma=\frac{g^2\left[1+2n_{b}(\omega_{0})\right]}
{32{\pi}\sqrt{\frac{9}{4}H^2-m^{*2}_{\Phi}}}\; \;  ; \; \omega_0 = 
-is^+_0. 
\label{gammasmall}
\end{equation} 

We can also study how the ``decay'' of the inflaton into lighter scalars
modifies the ``slow-roll'' evolution such as would occur in new inflationary
models. Since we are assuming a small field amplitude for our pertubation
theory to make sense, we cannot treat chaotic inflationary scenarios in the
same way.  

For this we now set $m^{2}_{\Phi}<0$ and study the situation in which the
inflaton is ``rolling'' from the top of the potential hill, during the 
stage of
quasi-exponential expansion. In this situation the amplitude expansion is 
valid
when the zero mode is close to the origin and at early times.

In this case it is convenient to write $m^2_{\Phi}=-\mu^2$. Now the poles are
on the real axis at the positions:
\begin{eqnarray}
s^{\pm}&=&\frac{-3H}{2}{\pm}\sqrt{\left(\frac{3H}{2}\right)^2+\mu^{2}-
\frac{g^2}{16{\pi}^2}\tilde{\Sigma}(s^{\pm})} \\
& \approx & s^{\pm}_0 \mp \frac{g^2}{16{\pi}^2}
\frac{\tilde{\Sigma}(s^{\pm})}{\sqrt{\left(\frac{3H}{2}\right)^2+\mu^{2}}}\\
s^{\pm}_{0}&=&\frac{-3H}{2}{\pm}{\sqrt{\left(\frac{3H}{2}\right)^2+\mu^{2}}}
\end{eqnarray}

For $g=0$ the pole that is on the positive real axis, $s^+_0$, is the one 
that
dominates the evolution of the inflaton down the potential hill (the growing
mode). Since $s^+_0 >0$ we find that
\begin{equation}
\tilde{\Sigma}(s^+_0) = \sum_2^{\infty}\frac{s^+_0}{n(s^+_0+nH)} >0 ,
\end{equation}
and we conclude that, to this order, the pole on the positive real axis is
shifted towards the origin. Therefore the rate of growth of the growing 
mode is
diminished by the decay into lighter scalars and the ``rolling'' is slowed
down. This is physically reasonable, since the ``decay'' results in a 
transfer
of energy from the inflaton to the massless scalars and the rolling of the
field down the potential hill is therefore slowed down by this decay process.
  
\subsection{Minimally coupled massless $\sigma$:}
Now we turn to the $\xi_{\sigma}=0$ case. The Green's functions are given by
(\ref{ggreater},\ref{glesser}) with the Hankel functions (\ref{minimalcoup}).
There are many features in common with the previous case: the kernel is
translationally invariant in time and periodic in imaginary time with
periodicity $\beta_H$. Thus it can again be expanded in terms of Matsubara
frequencies and the Laplace transform carried out in a straightforward 
manner.
However, for minimal coupling we find a new logarithmic infrared 
divergence in
the kernel along with the ultraviolet logarithmic singularity of the
conformally coupled case. Therefore, the $k$-integral in the kernel must be
performed with an infrared cutoff $\mu$. The logarithmic ultraviolet 
divergence
is handled in as before, and after this subtraction we find the 
self-energy to
be:
\begin{eqnarray}
\tilde{\Sigma}(s)=&&\overbrace{\sum^{\infty}_{1}H(\frac{1}{nH}-
\frac{1}{s+nH})-\frac{s}{s+H}}^{\xi_{\sigma}=1/6}-
\frac{4H}{3}\sum^{\infty}_{1}\frac{1}{n}(\frac{1}{s+nH}
-\frac{1}{s+3H+nH})\\\nonumber
\\\nonumber
&&+(\frac{1}{s}-\frac{1}{s+3H})[\frac{4H}{3}\ln{2}-
2H(\frac{2}{3}\ln(H/\mu)+\frac{14}{9})]\\\nonumber 
\\\nonumber
&&+\frac{4H}{3}(\frac{1}{s+H}-\frac{1}{s+2H})+
\frac{4H^2}{3}[\frac{1}{s^2}-\frac{1}{(s+3H)^2}].\nonumber
\end{eqnarray}
We notice that the self-energy for the
conformally coupled case is contained in the above expression. It is also
interesting to note that the imaginary part of the self-energy, on-shell,
receives no contribution from the infrared divergent piece, which only
contributes to a further renormalization of the ``pole mass''
$m^{*2}_{\Phi}$. We now focus on the case $4m^2_{\Phi}-9H^2>0$.

The expression for the decay rate is obtained by evaluating the imaginary 
part
of the self-energy at the pole $s^+_0$ (see eq. (\ref{dampin}) and found 
to be
\begin{equation} 
\Gamma= \overbrace{ \frac{g^2}{32\pi}[
 \frac{ \left[1+2 n_b(\omega_0) \right] }{M_{\Phi}}  }^{\xi_{\sigma}=1/6}
\left(1+\frac{4H^2}{m^{*2}_{\Phi}}\right)+\frac{8H^3}{{\pi}m^{*4}_{\Phi}}].
\end{equation}
with (\ref{bosefactor}).
The qualitative behaviour of the decay rate as a function of $H$ is 
similar 
to that
 in the conformally coupled case. We see that,
$\Gamma({\xi_{\sigma}}=0)>\Gamma({\xi_{\sigma}}=1/6)>\Gamma(\text{Minkowski})$.

\subsection{Analytic structure of the self-energy}

As mentioned earlier, the analytic structure of the self-energy in the De
Sitter background is drastically different from what one would expect in flat
space-time. We find that the self-energy for the case where $\Phi$ is 
unstable
does {\em not} display cuts in the $s$-plane. Though the reason for this 
strange
behaviour is not clear at present, we can try to understand it by taking the
flat-space limit i.e. the $H\rightarrow0$ limit. In this limit, it 
suffices to
look at $\Sigma(s)$ for $\xi_{\sigma}=1/6$. It is easily shown that the
additional terms in (3.25) are vanishingly small in the $H\rightarrow0$
limit. We thus obtain,
\begin{eqnarray}
\lim_{H\rightarrow0}\Sigma(s)=\lim_{H\rightarrow0}
[\sum^{\infty}_{1}&&H(\frac{1}{nH}-\frac{1}{s+nH})-\frac{s}{s+H}]\nonumber\\
\\\nonumber 
&&=\int^{\infty}_{H}(\frac{1}{x}-\frac{1}{s+x})dx-1\nonumber\\
\\\nonumber
&&=\ln(s/m)+\ln(m/H)-1\nonumber
\end{eqnarray}
where $m$ is a mass scale that serves as an infrared regulator. Hence,
\begin{eqnarray}
\tilde{\phi}(s)=\frac{s\phi_i}{s^2+(m^{2}_{\Phi}+
\frac{g^2}{16{\pi}^2}
\ln(m/H)-1)+\frac{g^2}{16{\pi}^2}\ln(s/m)}.
\end{eqnarray}

The above expression has a cut which can be chosen to run from 0 to 
$-\infty$,
so that it is to the left of the Bromwich contour. The discrete singularities
of the self-energy at $s=-nH$ have merged into a continuum in the small $H$
limit to give a cut along the negative real axis in the s-plane. In addition,
there are two simple poles in the first Riemann sheet at 
${\pm}im_{\Phi}+{\cal
O}(g^2)$.  Inverting the transform yields,
\begin{eqnarray}
\phi(t)&=&Z\phi_ie^{-{\Gamma}t/2}\cos(m_{\Phi}t+\alpha) \nonumber \\
&-&\phi_i\frac{g^2}{16{\pi}^2}{\int}_{0}^{\infty}d{\omega}
\frac{{\omega}e^{-{\omega}t}}{({\omega}^2+m_{\Phi}^2
+\frac{g^2}{16{\pi}^2}{\ln}({\omega}/m))^2+(\frac{g^2}{16{\pi}})^2}.
\label{nocut}
\end{eqnarray}
We see here, a Breit-Wigner form plus contributions from across the cut.

A flat-space analysis with $m_{\sigma}\neq 0$, carried out in a previous work
\cite{boya} reveals a very different picture. In particular, we find two cuts
extending from $s=\pm 2im_{\sigma}$ to ${\pm}i\infty$ and two simple poles
which move off into the second Riemann sheet above the two-particle threshold
when $m_{\Phi}>2m_{\sigma}$. However, upon inverting the transform and taking
$m_{\sigma}\rightarrow0$, we obtain
\begin{eqnarray}
\phi(t)=\phi_i\frac{g^2}{16{\pi}^2}{\int}_{0}^{\infty}d{\omega}
\frac{{\omega}\cos{{\omega}t}}{({\omega}^2-m_{\Phi}^2-\frac{g^2}
{16{\pi}^2}{\ln}({\omega}/m))^2+(\frac{g^2}{32{\pi}})^2}\label{yescut}.
\end{eqnarray}
A deformation of the contour of integration shows that eqs.(\ref{nocut}) and
(\ref{yescut}) are in fact identical and hence we have consistency (see the
Appendix).  Although the limits $H \rightarrow 0$ and $m_{\sigma} \rightarrow
0$ do not commute insofar as the analytic structure in the $s$-plane is
concerned, the time evolution obtained from the inverse Laplace transform is
unambiguously the same.

\section{The O(2) model and Goldstone Bosons}

While the inflaton is typically taken to be a singlet of any gauge or global
symmetry, there is no reason that it could not transform non-trivially 
under a
continuous global symmetry. If this occurs, and the symmetry is broken
spontaneously, the interesting possibility arises of dissipation of 
energy into
Goldstone bosons. Even if the inflaton is a singlet, one could ask about the
evolution of fields that belong to multiplets of a global continous
symmetry, during inflation. Our primary motivation, however, is to study
dissipative processes of 
the inflaton via the decay into Goldstone bosons in De Sitter space-time as
new non-equilibrium mechanisms.

The effect may be studied in the O(2) linear sigma model, in which the 
inflaton
is part of the O(2) doublet, with Lagrangian density
\begin{eqnarray}
{\cal L} =&&\frac{1}{2} [e^{3Ht}{\dot{\Phi}}^{2}-e^{Ht}
{(\vec{\nabla}{\Phi})}^2+e^{3Ht}{\dot{\pi}}^{2}-e^{Ht}
{(\vec{\nabla}{\pi})}^2]\\
&&+\frac{1}{2}e^{3Ht}({\mu}^2-12{\xi}H^2)({\Phi}^{2}+{\pi}^{2})-e^{3Ht}
\frac{\lambda}{4!}({\Phi}^{2}+{\pi}^{2})^2 +h(t)\Phi.\nonumber
\end{eqnarray}
Assuming that ${\mu}^2>12{\xi}H^2$, in the symmetry broken phase the field
$\Phi$ acquires a VEV
\begin{equation}
\langle\Phi\rangle=\sqrt{\frac{6({\mu}^2-12{\xi}H^2)}{\lambda}},
\end{equation}
and the Goldstone bosons are left with no mass terms, and hence are massless,
minimally coupled fields in De Sitter space-time\cite{goldstonebosons}; 
this is a fairly well
known result.

We now write 
\begin{equation}
\Phi(x,t)=\sqrt{\frac{6({\mu}^2-12{\xi}H^2)}{\lambda}}+{\phi}(t)+{\psi}(x,t)
\end{equation}
and use the tadpole method to impose:
\begin{eqnarray}
\langle\psi(x,t)\rangle=0,\\\nonumber
\langle\pi(x,t)\rangle=0.
\end{eqnarray}
Using eq.(4.3) it is easy to see that couplings of the form 
$\psi{\pi}^2$, are
automatically induced in the Lagrangian.  Carrying out the same analysis as
before and keeping only terms up to first order in $\phi(t)$, we find the
following equation of motion
\begin{eqnarray}
\ddot{\phi}+3H\dot{\phi}+m^2(t)\phi-i\frac{3{\lambda}
{\mu}^2}{2{\pi}^2}\int^{t}_{0}d{\tau}[K_{\phi}(t,\tau)+
\frac{1}{9}K_{\pi}(t,\tau)]\phi(\tau)=\tilde{h}(t),\\\nonumber
\text{where, } K(t,\tau)={\int}^{\infty}_{0}{k^2}dk
[G^{>2}_{k}(\tau,t)-G^{<2}_{k}(\tau,t)]e^{3H\tau}.
\end{eqnarray}

Here $m^2(t)$ is given by the tree-level mass term $2\mu^2$ plus 
contributions
proportional to the tadpole $\langle\pi^2\rangle(t)$. Since the Goldstone 
boson
is 
always minimally coupled to the curvature $\langle\pi^2\rangle(t)$ will grow
linearly with time and result in a non-trivial time-dependence for the mass
of the massive mode $\phi$. In this case the methods of the earlier 
section are
not applicable, and the definition of a ``decay rate'' will be plagued by
time-dependent ambiguities. 

Such a time-dependence will, however, be absent in a De Sitter space 
that exists forever, in which case the contribution from the tadpole will
simply be a divergence that is renormalised through a redefinition of the 
mass.
In such a situation, the contribution to the decay rate from the massive mode
would be subdominant 
due to 
the kinematics and the Goldstone modes would provide the dominant 
contribution
to the decay rate. Thus, neglecting $K_{\phi}(t,\tau)$ the equation of motion
has exactly the same form as eq.(3.5). The subsequent analysis and results
would 
be identical to those
obtained above in the section on minimally coupled fields.

\section{The Langevin equation}

Our results allow us to make contact with the issue of decoherence and the
stochastic description of inflationary 
cosmology\cite{staro,lang,habib,habib2}.

Decoherence is a fundamental aspect of dissipational dynamics and in the
description of non-equilibrium processes in the early
universe\cite{lang,habib,habib2}. The relationship between fluctuation 
and dissipation
as well as a stochastic approach to the dynamics of inflation can be explored
by means of the Langevin equation.

This section is devoted to obtaining the corresponding Langevin equation for
the non-equilibrium expectation value of the inflaton field (zero mode) 
in the
one loop approximation within the model addressed in this article.

The first step in deriving a Langevin equation is to determine the 
`system' and
`bath' variables, and subsequently integrate out the `bath' variables to 
obtain
an influence functional for the `system' degrees of freedom\cite 
{lang,habib,habib2}.

We first separate out the expectation value and impose the tadpole conditions
as follows:
\begin{equation}
{\Phi}^{\pm}={\phi}^{\pm}+{\psi}^{\pm}\hspace{0.1in};\hspace{0.1in}
\langle{\Phi}^{\pm}(\vec{x},t)\rangle={\phi}^{\pm}(t)\hspace{0.1in};
\hspace{0.1in}\langle{\sigma}^{\pm}(\vec{x},t)\rangle=0.
\end{equation}
The non-equilibrium effective action is defined in terms of the Lagrangian
given by equation(\ref{lagrangian}) as,
\begin{eqnarray}
e^{iS_{eff}}=\int{\cal{D}}{\psi}^{+}{\cal{D}}{\psi}^{-}{\cal{D}}{\sigma}^{+}
{\cal{D}}{\sigma}^{-}
e^{i\{S_{0}[{\phi}^{+}]-S_{0}[{\phi}^{-}]\}}e^{i\{S_{0}[{\psi}^{+}]-S_{0}
[{\psi}^{-}]\}}
\nonumber 
\\\nonumber
e^{i\{S_{0}[{\sigma}^{+}]-S_{0}[{\sigma}^{-}]\}}
e^{i\{S_{int}[{\psi}^{+},{\phi}^{+},{\sigma}^{+}]
-S_{int}[{\psi}^{-},{\phi}^{-},{\sigma}^{-}]\}}
\end{eqnarray}
where $S_{0}$ represents the action for free fields, and $S_{int}$ 
contains all
the remaining parts of the action. The effective action for the zero 
modes is
obtained by tracing out all the degrees of freedom corresponding to the
non-zero modes (or the `bath' variables), in a consistent loop 
expansion.  It
is useful to introduce the center of mass $(\phi(t))$ and relative $(R(t))$
coordinates as
\begin{equation}
\phi^{\pm}(t)=\phi(t)\pm\frac{R(t)}{2}
\end{equation}

Remembering the definitions of the Green's functions in eqs.(\ref{gplpl} -
\ref{gplmin}), we expand $\exp(iS_{int})$ up to order $g^2$ and consistently
impose the tadpole condition to obtain the effective action per unit volume
(because the expectation value is taken to be translationally invariant),
\begin{eqnarray}
\frac{S_{eff}}{\Omega}&=&{\int}dt
\left({\cal{L}}_0(\phi^+)-{\cal{L}}_0(\phi^-)\right)\\\nonumber
&-&\frac{ig^2}{16\pi^3}{\int}^{\infty}_{-\infty}dt{\int}^{t}_{-\infty}dt^{\prime}
{\int}d^3k 
a^3(t)a^3(t^{\prime})R(t)\phi(t^{\prime})
\left[G^{>2}_{k}(t,t^{\prime})-G^{<2}_{k}(t,t^{\prime})\right]\\\nonumber
&-&\frac{ig^2}{32\pi^3}{\int}^{\infty}_{-\infty}dt{\int}^{\infty}_{-\infty}dt^{\prime}
{\int}d^3k
a^3(t)a^3(t^{\prime})R(t)R(t^{\prime})
\left[G^{>2}_{k}(t,t^{\prime})+G^{<2}_{k}(t,t^{\prime})\right]\\\nonumber
&+& \tilde{h}(t)R(t) + \text{terms independent of $\phi$}\\\nonumber
\end{eqnarray}

As before $\tilde{h}(t)$ results from subtracting the
$\phi$-independent contribution of the tadpole from $h(t)$.

Now, it is not difficult to verify that the non-equilibrium Green's functions
given by eqn.(\ref{gfscale}, \ref{definef}) and obeying the boundary
conditions (\ref{bc}) have the property,
\begin{equation}
G^{>}_{k}(t,t^{\prime})=\left[G^{<}_{k}(t,t^{\prime})\right]^*. 
\end{equation}
This property guarantees that the second term in the effective action is 
real,
while the third term is pure imaginary. The imaginary, non-local, acausal 
part 
of the effective action gives a contribution to the path-integral that may
be written in terms of a stochastic field as
\begin{eqnarray}
\exp[&-&\frac{1}{2}{\int}dt{\int}dt^{\prime}
R(t){\cal{K}}(t,t^{\prime})R(t^{\prime})]{\propto}{\int}{\cal{D}}\xi
{\cal{P}}[\xi]
\exp[i{\int}dt\xi(t)R(t)]\\\nonumber
\cal{P}[\xi]&=&\exp[-\frac{1}{2}{\int}dt
{\int}dt^{\prime}{\xi}(t){\cal{K}}^{-1}(t,t^{\prime})
{\xi}(t^{\prime})]
\end{eqnarray}
with
\begin{eqnarray}
{\cal{K}}(t,t^{\prime})=-\frac{g^2}{16\pi^3}{\int}d^3k
a^3(t)a^3(t^{\prime})
[G^{>2}_{k}(t,t^{\prime})+G^{<2}_{k}(t,t^{\prime})]
\end{eqnarray} 
The non-equilibrium path integral now becomes
\begin{eqnarray}
Z{\propto}{\int}{\cal{D}}\phi{\cal{D}}R{\cal{D}}\xi{\cal{P}}[\xi]
\exp{i[S_{reff}(\phi,R)+{\int}dt\xi(t)R(t)]}
\end{eqnarray}
with $S_{reff}[\phi,R]$ being the real part of the effective action and with
$\cal{P}[\xi]$ the gaussian probability distribution for the stochastic noise
variable.

The Langevin equation is obtained via the saddle point condition\cite{lang}
\begin{equation}
\frac{{\delta}S_{reff}}{{\delta}R(t)}|_{R=0}=\xi(t)
\end{equation}
leading to
\begin{eqnarray}
\ddot{\phi}+3H\dot{\phi}+m_{\Phi}^2\phi
+i\frac{g^2}{16{\pi}^3}\int^{t}_{-\infty}dt^{\prime}{\int}d^{3}ka^{3}(t^{\prime})
[G^{>2}_{k}(t,t^{\prime})-G^{<2}_{k}(t,t^{\prime})]\phi(t^{\prime}) 
-\tilde{h}
(t)=
\frac{\xi(t)}{a^{3}(t)} \label{stochastic}
\end{eqnarray}
where the stochastic noise variable $\xi(t)$ has gaussian, but colored
correlations: 
\begin{eqnarray}
<<\xi(t)>>=0; \; \; \; <<\xi(t)\xi(t^{\prime})>>={\cal{K}}(t,t^{\prime})
\label{noisecorr}
\end{eqnarray}
The double brackets stand for averages with respect to the gaussian 
probability
distribution ${\cal{P}}[\xi]$. The noise is coloured and additive. Due to the
properties of the Green's functions (5.4), the noise kernel (which 
contributes
to the imaginary part of the effective action) is simply the $real$ 
$part$ of
$G^{>2}_{k}(t,t^{\prime})$. The dissipative kernel (which gives rise to the
non-local term in the Langevin equation) on the other hand is given by the
$imaginary$ $part$ of $G^{>2}_{k}(t,t^{\prime})$.  Consequently, the
$fluctuation-dissipation$ $theorem$ is revealed in the $Hilbert$ $transform$
relationship between the real and imaginary parts of the analytic function
$G^{>2}_{k}(t,t^{\prime})$.  The equation of motion (\ref{eqnofmotion}) 
is now
recognized to result from (\ref{stochastic}) by taking the average over the
noise.

One could now obtain the corresponding Fokker-Planck equation that describes
the evolution of the probability distribution function for 
$\phi$\cite{staro,fokker}.

We must emphasize here, that all the results discussed so far in this 
section,
are independent of the specifics of the FRW background spacetime, the 
masses of
the fields, and the temperature of the initial thermal state. The 
importance of
the Langevin equation resides at the fundamental level in that it 
provides a
direct link between fluctuation and dissipation including all the memory
effects and multiplicative aspects of the noise correlation functions. 

In particular in stochastic inflationary models it is typically
assumed \cite{staro} that the noise term is gaussian and white 
(uncorrelated).
This simplified stochastic description, in terms of gaussian
white noise, leads 
to a scale invariant spectrum of scalar density
perturbations\cite{staro}. Although this description is rather compelling,
within the approximations made in our analysis we see that for a
$\sigma$ field with an arbitrary mass and arbitrary couplings to the 
curvature
there is no
regime in which the correlations of the noise term
(\ref{noisecorr}) can be described by a Markovian delta function in
time\cite{lang}. 
One could speculate that some other couplings or higher order effects, or
perhaps some peculiar initial states could lead to a gaussian white 
noise, but
then one concludes that gaussian white noise correlations are by no means a
generic feature of the microscopic field theory.

We can speculate that our result of gaussian but correlated noise could have
implications 
for stochastic inflation. In particular it {\em may} lead to departures 
from a
scale 
invariant (Harrison-Zel'dovich) spectrum of primordial scalar density
perturbations, that can now be calculated within a particular microscopic 
model
using  
the non-equilibrium field-theoretical tools described in this work. Of course
this 
requires further and deeper study, which is beyond the scope of this article.

\section{Conclusions and further questions}

In this article we have studied the decay of inflatons into massless 
scalars in
De Sitter space. The motivation was to understand the non-equilibrium
mechanisms of inflaton relaxation during the stage of quasi-inflationary
expansion in early universe cosmology, in old, new and stochastic 
inflationary
scenarios. Chaotic inflation models cannot be treated within our 
approximation
scheme since the field will typically start at values that are large 
enough so
that the amplitude expansion is not valid. Having recognized the inherent
difficulties in defining a decay rate in a time dependent background, we have
given a practical definition that requires understanding the real-time
evolution of the inflaton interacting with lighter fields.  This led us 
to a
model in which an inflaton interacts with massless scalars via a trilinear
coupling, allowing for both minimal and conformal coupling to the curvature.

We obtained a decay rate at one loop order that displays a remarkable 
property.
With some minor modifications it can be interpreted as the stimulated 
decay of
the inflaton in a {\em thermal bath} at the Hawking temperature. This decay
rate is larger than that of Minkowski space because of the Bose-enhancement
factors associated with the Hawking temperature. The decay rate of the
minimally coupled case is larger than that of the conformal case which in 
turn
is larger than the Minkowski rate. We have shown explicitly that in the 
case of
new inflation, the dissipation of the inflaton energy associated with the 
decay
into conformally coupled massless scalars slows down further the rolling 
of the
inflaton down the 
potential hill.
We have also studied decay into Goldstone bosons.

To establish contact with the stochastic inflationary scenario, we 
derived the
Langevin equation for the coarse-grained expectation value of the inflaton
field to one loop order.  We find that this stochastic equation has a 
gaussian
but correlated (colored) noise. The two point correlation function of the 
noise
and the dissipative kernel fulfill a generalized fluctuation-dissipation
relation.

There are several potentially relevant implications of our results for 
old, new
and stochastic inflation. In the case of old inflation we see that 
dissipative
processes in the metastable maximum can contribute substantially to the
``equilibration'' of the inflaton oscillations, and more so because of the
enhanced stimulated decay for large Hubble constant. In the case of new
inflation, we have seen that these dissipative effects help slow the 
rolling of
the inflaton field down the potential hill, possibly extending the stage of
exponential expansion. 

Within the context of stochastic inflation, our results point out to the
possibility of incorporating deviations from a scale invariant spectrum of
primordial scalar density perturbations by the noise correlations, which
manifest the underlying microscopic correlations of the field theory. 
This is a
possibility that is worth exploring further.

Our results also point to further interesting questions: 
\begin{enumerate}
\item{Is it possible to
understand at a more fundamental level the connection between the decay rate
and the Hawking temperature of a bath {\em in equilibrium}? In 
particular, is
this connection maintained at higher orders?}

\item{Is it possible to relate the ``decay rate'' to the rate of particle
production via the interaction? Such a relation in Minkowski space-time 
is a
consequence of the existence of a spectral representation for the self-energy
but such a representation is not available in a time dependent background.}
\end{enumerate}

Answers to these questions will undoubtedly offer a much needed deeper
understanding of non-equilibrium processes in inflationary cosmology.

\acknowledgements

D.B would like to thank H. J. de Vega for
useful references and illuminating discussions, and NSF for support through
grant award No: PHY-9302534. D. B. and S. P. K. would like to thank Salman
Habib for 
illuminating conversations and comments.

S. Prem Kumar and R.Holman were partially supported by the U.S. Dept. of 
Energy
under 
Contract DE-FG02-91-ER40682.

\section{Appendix} 

In this appendix, we show explicitly, that (\ref{nocut}) and 
(\ref{yescut}) are
in fact identical expressions. The behaviour of $\phi(t)$ in Minkowski (given
by (3.57)) was obtained in reference\cite{boya} where $m_{\sigma}$ was 
allowed
to have non-zero values. Now, the results from this reference must obviously
coincide with the flat-space limits $(H \rightarrow 0)$ derived in this
paper. To see this, consider eqn.(\ref{yescut}):
\begin{eqnarray}
\phi(t)&=&\phi_i\frac{g^2}{16{\pi}^2}{\int}_{0}^{\infty}d{\omega}
\frac{{\omega}\cos{{\omega}t}}{({\omega}^2-m_{\Phi}^2-\frac{g^2}
{16{\pi}^2}{\ln}({\omega}/m))^2+(\frac{g^2}{32{\pi}})^2}\nonumber
\\\nonumber
\\\nonumber
&=&\frac{\phi_i}{\pi}Re[{\int}_{0}^{\infty}-id{\omega}
({\omega}e^{{i\omega}t})[\frac{1}{{\omega}^2-m_{\Phi}^2-\frac{g^2}
{16{\pi}^2}{\ln}({\omega}/m)-i\frac{g^2}{32{\pi}}}\nonumber
\\\nonumber
&-&\frac{1}{{\omega}^2-m_{\Phi}^2-\frac{g^2}
{16{\pi}^2}{\ln}({\omega}/m)+i\frac{g^2}{32{\pi}}}]].\nonumber
\end{eqnarray}
The branch cut due to the logarithm in the denominator runs from $0$ to
$+\infty$, and the contour of integration may be chosen to run from $0$ to
$+\infty$ on the first sheet. It is easy to see that the integrand has 4 
simple
poles, one in each quadrant. In particular, the pole in the first 
quadrant is
at
\begin{equation}
{\omega}_1=m_\Phi+\frac{g^2}{32{\pi}^2m_\Phi}
{\ln}(m_{\Phi}/m)+i\frac{g^2}{64{\pi}m_\Phi}.\nonumber
\end{equation}

Furthermore, the exponential is well behaved at infinity in the first
quadrant. We can therefore deform the contour of integration so that it runs
>from $0$ to $+i\infty$, picking up the residue from one simple pole.  
Writing
$\omega=iz$, and noticing that the logarithm picks up an imaginary part 
${i\pi}/2$
we now have,
\begin{eqnarray}
\phi(t)&=&\frac{\phi_i}{\pi}Re[{\int}_{0}^{\infty}dz
(ize^{-zt})[\frac{1}{-z^2-m_{\Phi}^2-\frac{g^2}
{16{\pi}^2}{\ln}(z/m)-i\frac{g^2}{16{\pi}}}\nonumber
\\\nonumber
\\\nonumber
&-&\frac{1}{-z^2-m_{\Phi}^2-\frac{g^2}
{16{\pi}^2}{\ln}(z/m)}]+ {\rm Res}[\frac{{-i\omega}e^{{i\omega}t}}
{{\omega}^2-m_{\Phi}^2-\frac{g^2}
{16{\pi}^2}{\ln}({\omega}/m)-i\frac{g^2}{32{\pi}}}]|_{\omega={\omega}_1}].
\nonumber
\\\nonumber
\\\nonumber
\\\nonumber
\\\nonumber
&=&\frac{\phi_i}{\pi}[-\frac{g^2}{16{\pi}}{\int}_{0}^{\infty}dz
\frac{ze^{-zt}}{(z^2+m_{\Phi}^2+\frac{g^2}
{16{\pi}^2}{\ln}(z/m))^2+(\frac{g^2}{16{\pi}})^2}\nonumber
\\\nonumber
\\\nonumber
&+&2{\pi}Re[\lim_{{\omega}\rightarrow{\omega_1}}(\frac{{\omega}e^{{i\omega}t}
({\omega}-{\omega}_1)}
{{\omega}^2-m_{\Phi}^2-\frac{g^2}
{16{\pi}^2}{\ln}({\omega}/m)-i\frac{g^2}{32{\pi}}})]]\nonumber
\end{eqnarray}
Absorbing the real part of the self-energy on-shell as an additional mass
renormalization, and going through the subsequent algebra we obtain our 
result,
\begin{eqnarray}
\phi(t)={\phi}(0)(1+\frac{g^2}{32{\pi}^2{m_\Phi}^2})e^{-\frac{g^2t}{64{\pi}
{m_\Phi}}}\cos(m_{\Phi}t)\nonumber
\\\nonumber
-\phi_i\frac{g^2}{16{\pi}^2}{\int}_{0}^{\infty}dz
\frac{ze^{-zt}}{(z^2+m_{\Phi}^2+\frac{g^2}
{16{\pi}^2}{\ln}(z/m))^2+(\frac{g^2}{16{\pi}})^2}.\nonumber
\end{eqnarray}

\end{document}